\documentclass[journal=jacsat,manuscript=article]{achemso}

\usepackage[version=3]{mhchem} 

\author{Shaojun Wang}
\affiliation{Department of Applied Physics and Institute for Photonic Integration, Eindhoven University of Technology, P.O. Box 513, 5600 MB Eindhoven, The Netherlands}
\alsoaffiliation{MOE Key Laboratory of Modern Optical Technologies and Jiangsu Key Laboratory of Advanced Optical Manufacturing Technologies, School of Optoelectronic Science and Engineering, Soochow University, Suzhou 215006, China}
\email{swang.opto@suda.edu.cn}

\author{T.V. Raziman}
\affiliation{Department of Applied Physics and Institute for Photonic Integration, Eindhoven University of Technology, P.O. Box 513, 5600 MB Eindhoven, The Netherlands}

\author{Shunsuke Murai}
\affiliation{Department of Applied Physics and Institute for Photonic Integration, Eindhoven University of Technology, P.O. Box 513, 5600 MB Eindhoven, The Netherlands}
\altaffiliation{Department of Material Chemistry, Graduate School of Engineering, Kyoto University, Nishikyo-ku,
Kyoto 615-8510, Japan}

\author{Gabriel W. Castellanos}
\affiliation{Department of Applied Physics and Institute for Photonic Integration, Eindhoven University of Technology, P.O. Box 513, 5600 MB Eindhoven, The Netherlands}

\author{Ping Bai}
\affiliation{Department of Applied Physics and Institute for Photonic Integration, Eindhoven University of Technology, P.O. Box 513, 5600 MB Eindhoven, The Netherlands}

\author{Anton Matthijs Berghuis}
\affiliation{Department of Applied Physics and Institute for Photonic Integration, Eindhoven University of Technology, P.O. Box 513, 5600 MB Eindhoven, The Netherlands}
\author{Rasmus H. Godiksen}
\affiliation{Department of Applied Physics and Institute for Photonic Integration, Eindhoven University of Technology, P.O. Box 513, 5600 MB Eindhoven, The Netherlands}
\author{Alberto G. Curto}
\affiliation{Department of Applied Physics and Institute for Photonic Integration, Eindhoven University of Technology, P.O. Box 513, 5600 MB Eindhoven, The Netherlands}
\author{Jaime Gómez Rivas}
\affiliation{Department of Applied Physics and Institute for Photonic Integration, Eindhoven University of Technology, P.O. Box 513, 5600 MB Eindhoven, The Netherlands}
\email{j.gomez.rivas@tue.nl}

\title
 {Collective Mie Exciton-Polaritons in an Atomically Thin Semiconductor}

\abbreviations{IR,NMR,UV}

\begin{document}


\begin{abstract}
Optically induced Mie resonances in dielectric nanoantennas feature low dissipative losses and large resonant enhancement of both electric and magnetic fields. They offer an alternative platform to plasmonic resonances to study light-matter interactions from the weak to the strong coupling regimes. Here, we experimentally demonstrate the strong coupling of bright excitons in monolayer WS$_2$ with Mie surface lattice resonances (Mie-SLRs). We resolve both electric and magnetic Mie-SLRs of a Si nanoparticle array in angular dispersion measurements. At the zero detuning condition, the dispersion of electric Mie-SLRs (e-SLRs) exhibits a clear anti-crossing and a Rabi-splitting of 32 meV between the upper and lower polariton bands. The magnetic Mie-SLRs (m-SLRs) nearly cross the energy band of excitons. These results suggest that the field of m-SLRs is dominated by out-of-plane components that do not efficiently couple with the in-plane excitonic dipoles of the monolayer WS$_2$. In contrast, e-SLRs in dielectric nanoparticle arrays with relatively high quality factors (Q $\sim$ 120) facilitate the formation of collective Mie exciton-polaritons, and may allow the development of novel polaritonic devices which can tailor the optoelectronic properties of  atomically thin two-dimensional semiconductors.     

\end{abstract}
\newpage

\section{Introduction}

Mie resonances have recently triggered significant interest in the field of nanophotonics due to their unique optical properties,\cite{kuznetsov2016optically} which can lead to novel optical devices such as super-cavities,\cite{rybin2017optical} optical sensors,\cite{yesilkoy2019ultrasensitive} light-emitting metasurfaces,\cite{murai2020light,vaskin2019,bidault2019,sun2019electromagnetic} and lasers.\cite{ha2018directional} These resonances arise from displacement currents in dielectric or semiconducting nanoparticles with high refractive indices. Dielectric nanoparticles offer an alternative platform to plasmonic nanostructures for nanophotonics due to the low material absorption and the rich diversity of their electromagnetic modes.\cite{evlyukhin2010optical} For instance, the interference of electric and magnetic resonances in dielectric nanoantennas satisfying the Kerker condition leads to the full suppression of backward scattering.~\cite{kerker1983electromagnetic,staude2013tailoring,babicheva2017resonant,li2018engineering,cihan2018silicon} The combination of dielectric nanoparticles in arrays can lead to collective resonances with narrow line-widths.\cite{castellanos2019lattice} Symmetry protection for radiation losses in these arrays can generate bound states in the continuum with infinite quality factors.\cite{stillinger1975bound,marinica2008bound,kodigala2017lasing} 
The coupling of Mie resonances with excitons is expected to tailor the performance of optoelectronic materials.\cite{ebbesen2016hybrid}, e.g., two-dimensional (2D) semiconductors, and enable new phenomena such as polariton lasing and condensation.\cite{kena2010room,Plumhof_2013,Daskalakis_2014,ramezani2017plasmon} 

Atomically thin transition metal dichalcogenides (TMDs) such as MoS$_2$ or WS$_2$ are typical 2D semiconductors, exhibiting unique optoelectronic and structural properties. Monolayer TMDs display a direct band gap compared with bulk or multilayer TMDs.~\cite{mak2010atomically,splendiani2010emerging} The photogenerated bright excitons in monolayer TMDs possess very large (hundreds of meV) binding energies and are stable at room temperature.~\cite{mak2010atomically,splendiani2010emerging,chernikov2014exciton} Despite being atomically thin, monolayer TMDs show strong absorption in the visible and near infrared. Further enhancement and control of light–matter interactions is still possible through the integration of monolayers into optical architectures.

Different types of photonic or plasmonic nanostructures have been demonstrated to couple with monolayer TMDs.~\cite{lu2017nearly,ao2018unidirectional,bucher2019tailoring,wang2016coherent,zheng2017manipulating, wen2017room, zhang2018photonic,chervy2018room,wang2019limits,xie2020coherent} In the weak coupling regime, the external nanostructures enhance the absorption efficiency,~\cite{lu2017nearly} modify the local density of optical states through the Purcell effect,~\cite{sortino2019enhanced} and increase the directivity of the emission of TMDs.~\cite{ao2018unidirectional,bucher2019tailoring} When the coherent energy exchange between excitonic transitions in TMDs and the optical resonances in nanostructures is faster than any damping rate in the system, the interaction between them reaches the strong coupling regime and leads to the formation of hybrid light-matter states, i.e., exciton-polariton states. Very recently, theoretical works have proposed that magnetic resonances supported by single silicon nanoparticles can lead to Mie exciton-polaritons in monolayer TMDs.~\cite{tserkezis2018mie,lepeshov2018tunable,wang2019resonance} These works use a core-shell structure with the monolayer TMD onto Si spheres to maximize the coupling strength. However, in experiments it is only possible to deposit Si nanoparticles on top of flat TMDs.~\cite{lepeshov2018tunable,wang2019resonance} The weak electromagnetic field on the surface of the Si nanoparticle and the high radiative damping rates of Mie resonances hinder the interaction with the monolayer TMDs.     

In this letter, we successfully achieve strong coupling of bright excitons in monolayer TMDs with collective Mie resonances in periodic arrays of nanoparticles, i.e., Mie surface lattice resonances (Mie-SLRs). We use monolayer WS$_2$ coupled to low-loss polycrystalline Si nanodisk arrays embeded in an optically homogeneous medium. We resolve both electric and magnetic Mie-SLRs in the dispersion measurements with different polarizations. The atomically thin monolayer provides a simple tool to detect the intensity of the electromagnetic field supported by different optical modes.\cite{yesilkoy2019ultrasensitive} When we deposit the monolayer WS$_2$ on top of the array, the energy of electric Mie-SLRs (e-SLRs) redshifts 10 meV, while the magnetic Mie-SLRs (m-SLRs) shows a smaller redshift of only 2 meV. At the zero detuning condition between excitons and SLRs, the angular dispersion of e-SLRs exhibits a clear anti-crossing with a Rabi-splitting of 32 meV between the upper and the lower polariton bands. For the m-SLRs, the dispersion crosses the energy of excitons. The different coupling strength and Rabi-splitting of the e-SLRs and m-SLRs indicates that the electric field of m-SLRs is dominated by out-of-plane components that do not couple efficiently with the in-plane excitonic dipoles in monolayer TMDs. In contrast, e-SLRs in dielectric nanoparticle arrays with relatively high quality factors (Q $\sim$ 120) facilitate the formation of collective Mie exciton-polaritons. Finally, we use numerical simulations to compare Si and Ag nanodisk arrays with the same symmetry and lattice constant. Si arrays with sharper resonance and stronger near-field enhancement are more efficient for achieving strong coupling and forming exciton-polaritons in atomically thin TMDs. Our results enrich the nanophotonic tools for investigating the polariton physics in 2D semiconductors and pave the way for the design of novel polaritonic devices. 

\section{Methods}
\textbf{Sample fabrication}. The Si nanoparticle array was fabricated using electron-beam lithography and selective dry etching. Polycrystalline Si thin films were grown on a fused silica substrate by low-pressure chemical vapor deposition employing SiH$_4$ gas as the source of Si. A resist (NEB22A2, Sumitomo) was spin cast onto the Si film and patterned by electron-beam lithography and development. The Si film was vertically etched using a Bosch process with SF$_6$ and C$_4$H$_8$ gases, and the resist residue was etched away by oxygen dry etching. A high quality monolayer of WS$_2$ with a size of 30 x 70 $\mu m^2$ was mechanically exfoliated from a synthetic single crystal (HQ Graphene). The monolayer region of the flake was determined with white light microscopy and extinction measurements. The monolayer sample was exfoliated onto an optically transparent and flexible PDMS substrate. The monolayer on the PDMS was aligned under a microscope and softly transferred mostly onto the Si nanoparticle array and partially onto the flat quartz substrate for reference measurements.

\textbf{Optical extinction measurements of the bare array}. The optical extinction of the bare nanoparticle array (2 x 2 mm$^2$) covered with a 200 mm thick PDMS superstrate was measured with a collimated white light beam and with the sample mounted on a rotation stage to change the angle of incidence. The zero-order transmission spectrum was recorded with a fiber-coupled spectrometer (USB2000+, Ocean Optics). The extinction is defined as 1-$T/T_0$, where $T$ is the transmittance through the nanoparticle array and $T_0$ is the transmittance through the reference measured outside the array.

\textbf{Optical extinction measurements under a microscope}. The extinction spectra of the monolayer flake on the quartz substrate and the Si nanoparticle array were measured using an optical microscope. The samples were aligned along the optical axis of the microscope and illuminated with quasi-collimated white light. The light transmitted through the samples was collected using a microscope objective lens (Nikon CFI S Plan Fluor ELWD 20x, N.A. = 0.45), and imaged with a spectrometer (Princeton Instrument SpectraPro 300i) and an electron-multiplying charge-coupled device camera (Princeton Instruments ProEM: 512). The angular dependent extinction spectra of the nanoparticle array with/without the monolayer were recorded by rotating the sample.

\textbf{Coupled oscillators model fit}. The hybrid system can be approximately described with a model of two coupled harmonic oscillators:\begin {equation}\begin{bmatrix}\omega_{e-SLRs}-\frac{i\Gamma_{e-SLRs}}{2} & g \\g & \omega_{ex}-\frac{i\gamma_{ex}}{2} \end{bmatrix}\left(\begin{array}{c}\alpha\\ \beta\end{array}\right)=\omega\left(\begin{array}{c}\alpha\\ \beta\end{array}\right),\end {equation} where $\omega_{e-SLRs}$ and $\omega_{ex}$ are the resonant energies of the bare SLRs and excitons, respectively; $\Gamma_{e-SLRs}$ and $\gamma_{ex}$ represent the damping rates of e-SLRs and excitons, and $g$ is the coherent coupling strength. Diagonalizing the Hamitonian matrix yields the new exciton-polaritonic eigenvalues $\omega_{\pm}$, defining the energies of the UP and LP bands and the Hopfield coefficients $\alpha$ and $\beta$, the squares of which give the weight fractions of excitons and SLRs with $\mid\alpha\mid^2+\mid\beta\mid^2=1$. The value of the Rabi splitting, $\omega_{+}-\omega_{-}=\sqrt{4g^{2}-\frac{(\Gamma_{e-SLRs}-\gamma_{ex})^{2}}{4}}$, is given at the condition of zero detuning, namely, $\omega_{e-SLRs}=\omega_{ex}$. We extract $\omega_{e-SLRs}$ and the weight fractions assuming $\omega_{ex}$ = 2.016~eV and fitting the peak positions of the UP and LP bands to the model. 

\textbf{Simulations}. We use a high-accuracy surface integral method for periodic scatterers,~\cite{Gallinet_2010,Raziman_2015} and modelled the nanoparticles as cylinders (Height $H=90$ nm and diameter $D=126$ nm for silicon, $H=40$ nm and $D=74$ nm for silver) to obtain the best agreement with the experimental extinction. The size of the unit cell is 420 $\times $420 nm$^2$ in the $xy$-plane, with periodic boundary conditions applied to simulate an infinite array. The refractive index of the surrounding medium is set to 1.43. For the dielectric function of silicon, we used the measured values by Aspnes and Studna,~\cite{aspnes1983dielectric} with the imaginary part increased by 5 times to achieve the most accurate results compared to the measurements, as the etching process is likely to introduce imperfections on the particles that lead to an increased absorption. The permittivity of silver was taken from Palik,~\cite{palik_handbook_1998} and a conformal 5 nm thick alumina spacer layer (material parameters from Boidin~\cite{boidin_pulsed_2016}) was placed on top of the silver particle array to simulate experimental conditions.
\section{Results $\&$ Discussion}
\textbf{Mechanism and Sample Design.} Figures 1a and b illustrate the mechanism of in-plane strong light-matter coupling that leads to the formation of collective Mie exciton-polaritons. The resonant in-plane electric field associated with the nanoparticle array couples to the confined excitons in the 2D semiconductor on top of the array (Figure 1a). This coupling leads to hybridization and the formation of the lower (LP) and upper (UP) exciton-polaritons separated by the Rabi energy $\Omega_{R}$ (Figure 1b). To observe Mie exciton-polaritons, we fabricated a polycrystalline-Si nanoparticle array with a size of 2 $\times$ 2 mm$^2$ on a fused quartz substrate by electron-beam lithography and selective dry etching (see Methods for details). A scanning electron microscope image of the nanoparticle array is shown in Figure 1d. We softly transferred an exfoliated WS$_2$ flake on top of the array and left the flexible polydimethylsiloxane (PDMS) as a superstrate (Figure 1c).\cite{eizagirre2019preserving} Accordingly, the Si nanoparticle array and the flake were sandwiched in a nearly homogeneous dielectric environment to enhance the coherent in-plane scattering of light from the nanoparticles.\cite{de2007colloquium,auguie2008collective,giannini2010lighting} The array has a square symmetry with a lattice constant $P$ = 420 nm. The individual nanoparticles are nanodisks with a height of 90 nm and a diameter of 130 $\pm $ 4 nm. A bright-field microscope image of the nanoparticle array with the WS$_2$ flake partially on top is shown in Figure S1.

\textbf{Mie-SLRs of the Bare Array.} We first characterize the optical resonances of the bare Si nanoparticle array without the flake on top but with a PDMS superstrate. We illuminated the particle array sample with a collimated white light beam and measured its angular extinction dispersion (see Methods). The incident wave vector lies in the $yz$ plane and projects its parallel component $k_\parallel$ onto the surface of the array ($xy$ plane). We set the polarization of the incident beam along the $x$ ($y$)-axis as TE (TM) polarization. The dispersion curves of both TE and TM polarized modes (Figures 2a and b) exhibit a sharp resonance ($\sim$2.0~eV) and a broad one ($\sim$2.4~eV), which are evident at the normal incidence condition (Figure 2c). The broad extinction peak is associated with the localized Mie-resonances in individual nanoparticles and shows almost no dispersion. The wave vector dependence of the sharp resonance follows the condition of in-plane diffraction orders, i.e., Rayleigh anomalies (RAs). Interestingly, TE/TM polarization modes show both linear and parabolic dispersion curves, which are associated with the (1, 0), (-1, 0), and (0, $\pm 1$) order RAs. The results are different from previous studies of plasmonic nanoparticle arrays, in which linear and parabolic dispersion are selected by the different polarization due to the existence of only electric dipole resonances.\cite{zakharko2018radiative,wang2018rich,le2019enhanced} In addition, the sharp peak with a high quality factor (Q-factor) of 120 approximately, estimated by fitting it with a Lorentzian, displays a pronounced extinction as large as ~0.92 at the energy of 2.0~eV. This peak corresponds to the so-called Mie-SLRs, emerging from the coherent radiative coupling between the Mie scatters enhanced by the in-plane diffracted orders, i.e., the RAs.\cite{murai2020sc}

We analyse more in detail the extinction spectra around the peak of the Mie-SLR (Figures 2d-f). A weak extinction shoulder at the energy around 2.05~eV can be appreciated at normal incidence (Figures 2c and f). When increasing the in-plane wave vector, this weak extinction peak follows a parabolic (linear) dispersion for TE (TM) polarization. The opposite relation between the polarization and the dispersion of the main Mie-SLR peak and the weaker peak is due to the excitation of both electric and magnetic dipolar resonances in arrays of dielectric nanoparticles. The polarized incident beam generates an electric dipole in the nanoparticles oriented along the polarization direction, and the displacement current density generates an orthogonal magnetic dipole. The radiative coupling of electric or magnetic dipoles on the nanoparticles is enhanced by the orthogonal diffraction orders to form e- and m-SLRs, respectively.\cite{babicheva2017resonant,li2018engineering,castellanos2019lattice,murai2020sc} 

\textbf{Rabi Splitting of the Hybrid System.} We only need to consider the in-plane components of the electric field on the upper surface of the nanoparticle array for coupling with the in-plane excitonic dipoles of the TMD monolayer. We use surface integral equations (SIE, see Methods) to simulate the field enhancement at this surface with a plane wave illumination at normal incidence ($k_\parallel$ = 0). The in-plane field enhancement distribution at the energies of 2.006~eV (Figure 3a) and 2.036~eV (Figure 3b) are very similar, with vertical RA bands of enhanced field. This similar field distribution indicates that the in-plane field at these two energies originates from the same mode, namely the e-SLRs. As expected, the in-plane field enhancement is more pronounced at the peak position (2.006~eV) than at the edge (2.036~eV). The simulations of the total field (Figures 3d, S5-S7) indicate that out-of-plane field components are dominant for the m-SLRs.~\cite{castellanos2019lattice} In particular, we see a total field enhancement distribution at 2.006 eV (Figure 3c) very different to that at 2.036 eV (Figure 3d) with the field enhanced along horizontal bands. This different total field distribution indicates that the m-SLR dominates the total field at 2.036 eV, but with mainly a out-of-plane field. Therefore, the light-matter coupling with TMD monolayers is dominated by the in-plane field of the e-SLRs rather than by the out-of-plane field of m-SLRs. When the energy exchange between the excitonic transitions and the e-SLRs is faster than their damping rates, the regime of strong light-matter coupling is reached. The energy of excitons and e-SLRs splits into two new hybrid light-matter states, i.e., the UP and LP states, that we call Mie exciton-polaritons. The corresponding extinction spectra of the uncoupled and hybrid systems are shown in Figure 1f. The energy difference between UP and LP defines the Rabi-splitting $\Omega_{R}$ and is equal to twice the coupling strength $g$.

Next, we evaluate the strength of light-matter interactions with e- and m-SLRs from the extinction measurements. We record the extinction spectra of the nanoparticle array without (Figures 4a and c) and with (Figures 4b and d) the WS$_2$~1L on top by varying the angle of incidence from $\theta=0^{\circ}$ to $22^{\circ}$. For TM polarization, the (0, $\pm 1$) e-SLRs dominate the extinction spectra as indicated by the solid-red guide to the eye in Figure 4a. We note that the spectral full width at half-maximum ($\Gamma _{e-SLRs}$) increases for larger angles of incidence. This increase is similar to the one observed in plasmonic arrays and it is due to the reduction in detuning between the Mie-resonances and the RAs.\cite{le2019enhanced,murai2020sc} For the case of TE polarization, the ($\pm 1$, 0) e-SLRs of the bare array, which are degenerate at $\theta=0^{\circ}$, split into two bands (denoted by the black arrows in Figure 4c) as the angle of incidence increases. The m-SLR around the resonant energy of the excitons (vertical black dashed line in Figures 4a-d) has a weaker extinction as indicated by the solid-red guide to the eye in Figure 4c. Due to the large dispersion of the ($\pm 1$, 0) diffraction orders, we only investigate the (0, $\pm 1$) SLRs of both magnetic (TE) and electric (TM) modes. With WS$_2$~1L on top of the array, we find that the dispersion of the e-SLR splits into the LP and UP bands (the solid-blue guide to the eye in Figure 4b). Unlike the e-SLR, the m-SLR (solid-blue guide to the eye in Figure 4d) is similar to the bare nanoparticle array shown in Figure 4c (solid-red guide to the eye). The main difference between Figures 4c and d is that the extinction spectrum of the excitons is superimposed on that of the m-SLRs. These results indicate that strong coupling of excitons in WS$_2$~1L takes only place with e-SLRs and not with m-SLRs.

To quantify the coupling strength between Mie-SLRs and excitons, we analyze the angular dispersion of the hybrid system. We plot the energies of the extinction peaks as a function of the incident in-plane momentum $k_\parallel$ in Figures 5a and c. The blue triangles represent the high and low energy bands of the coupled system. The red squares correspond to the energies of the bare SLRs. In the case of m-SLRs (Figure 5c), the upper and lower energy bands follow the dispersion curves of the bare Mie-SLR and excitons, respectively. The energies of bare m-SLRs redshift 2 meV and overlap with the upper band. Therefore, we confirm that magnetic dipole resonances cannot strongly couple with bright excitons in monolayer TMDs. Instead, the electric dipole resonances are efficient to achieve strong coupling with 2D semiconductors. The two new bands in Figure 5a, corresponding to the UP and LP bands, exhibit a clear anti-crossing at the zero detuning between e-SLR and excitons. We fit the angular dispersion of the UP and LP bands to a model with two coupled harmonic oscillators as shown by the blue curves in Figure 5a (see Methods for details). We find that the band of uncoupled e-SLRs (red solid curve) redshifts 10 meV when the monolayer WS$_2$ is transferred on top of the nanoparticle array compared with the bare array (also see Figure S4b). We also obtain the Rabi energy $\Omega_{R}$ = 32 meV at $k_\parallel$= 2.4 $\mu m^{-1}$, when the UP or LP state is half mixed with e-SLRs and half with excitons (Figure 5a). Due to the high Q-factor (Q$\sim$120) of e-SLRs in the Si nanoparticle array, the Rabi energy satisfies the strong coupling condition, i.e., $\Omega_{R}>\gamma _{ex}, \Gamma _{e-SLRs} $, where $\gamma _{ex}$ = 25 meV\cite{wang2016coherent,chervy2018room,wang2019limits,eizagirre2019preserving}, $\Gamma _{e-SLRs} $ = 18 meV are the line widths of the extinction spectra of excitons and e-SLRs, respectively. 
These linewidths were estimated by Lorentzian fits to the spectra. The results demonstrate that Si nanoparticle arrays are a reasonable platform for the generation of collective Mie exciton-polaritons in an atomically thin semiconductor.   

\textbf{Comparison of the Mie and Plasmonic Array.} Combining these results with our previous investigations of strong light-matter coupling with plasmonic arrays\cite{wang2019limits,ramezani2017plasmon,berghuis2019enhanced,ramezani2019ultrafast}, we raise the question of whether Mie or plasmonic nanoparticle arrays are more efficient for achieving strong coupling with bright excitons in monolayer TMDs. We used simulations to obtain the answer. The coupling efficiency of Mie or plasmonic SLRs with in-plane excitonic dipoles is proportional to the in-plane field supported by SLRs. Hence, we first calculate the in-plane field enhancement of SLRs and define the ratio of coupling efficiency of Mie and plasmonic nanoparticle arrays. To enter the strong coupling regime, the Rabi-energy should be larger than the line widths of SLRs. Sharper resonances allow more cycles of Rabi oscillations and are easier to satisfy the criterion of strong coupling. We thus continue to evaluate the line widths of Mie and plasmonic SLRs. 

Considering the measurements, we simulate the extinction spectra and in-plane field enhancement factor of silicon and metallic nanoparticle arrays using SIE (details in Methods). We use the experimental values of the height of the Si nanoparticles (90 nm) and 40 nm for the Ag nanoparticles.\cite{wang2019limits,berghuis2019enhanced,ramezani2017plasmon,le2019enhanced} The lattice constant of the square array is the same for both Si and Ag arrays and equal to 420 nm. To find the same energy e-SLRs, we tune the diameter of the Si and Ag nanodisks to 126 nm and 74 nm, respectively. The extinction spectra at normal incidence and the root-mean-square value of the in-plane enhancement factor within a unit cell of the bare particle array are shown in Figures 6a and b, respectively. The simulated spectrum of Si array (blue curve in Figure 6a) qualitatively agrees with the measured result (Figure 2f). For comparison, we focus on the e-SLRs of both Si and Ag array. The line-width of the Si array ($\Gamma _{Si} \sim $ 20 meV) is narrower than that of the Ag array ( $\Gamma _{Ag} \sim $ 34 meV) due to the lower material absorption of Si comparing with Ag. The simulated damping rates of Si and Ag array are qualitatively comparable with the current ($\Gamma _{Si} \sim $ 18 meV) and previous experimental measurements in similar Ag arrays ($\Gamma _{Ag} \sim $ 43 meV) ,\cite{wang2019limits} respectively. The extinction of Si array is a factor of 1.35 larger than the extinction of the Ag array (Figure 6a). The narrower and higher extinction peak of Si array leads to a stronger in-plane near field enhancement (Figure 6b), suggesting that Si array is more efficient than Ag array to achieve strong coupling with monolayer TMDs.

\section{Conclusions}
In summary, we have demonstrated an alternative nanophotonic structure, Si nanoparticle arrays, for achieving strong light-matter coupling and collective Mie exciton-polaritons in monolayer WS$_2$. The in-plane electromagnetic field associated with e-SLRs in Si nanoparticle arrays allows to strongly couple to in-plane excitonic dipoles in monolayer TMDs. At room temperature, we observe Rabi splitting when the energy of e-SLR is tuned to the energy of the excitons. However, the orthogonality between the out-of-plane field distribution with the in-plane excitons prevents the formation of exciton-polaritons for the m-SLR. We would expect that the m-SLR can be applied to couple with out-of-plane dipolar emitters, e.g., direct-bandgap interlayer excitons in TMD heterostructures.\cite{paik2019interlayer} In addition, Si nanoparticle arrays benefit from lower absorption and stronger electromagnetic field enhancement compared to Ag nanoparticle arrays with similar dimensions. Our findings contribute to the understanding of light-matter interactions at the nanoscale and pave the way for the investigation and design of low-loss polaritonic devices.

\section{Associated Content}
The Supporting Information includes the microscope image of the sample, extinction spectra of the sample, electric field enhancement factor of bare Si nanoparticle array, and Ag nanoparticle array.

\section{Author Information}
Notes

The authors declare no competing financial interest.

\section{Acknowledgements}
The authors thank the Innovational Research Incentives Schemes of the Nederlandse Organisatie voor Wetenschappelijk Onderzoek (NWO) (Vici grant nr. 680-47-628 and Gravitation grant nr. 024.002.033), and  the Ministry of Education, Culture, Sports, Science and Technology (MEXT, Japan)(17KK0133, 19H02434) for financial support. S.Wang was supported by Priority Academic Program Development (PAPD) of Jiangsu Higher Education Institutions. Numerical simulations in this work were carried out on the Dutch national e-infrastructure with the support of SURF Cooperative.

\bibliography{achemso-demo}

\begin{figure}
\includegraphics[width=0.8\textwidth]{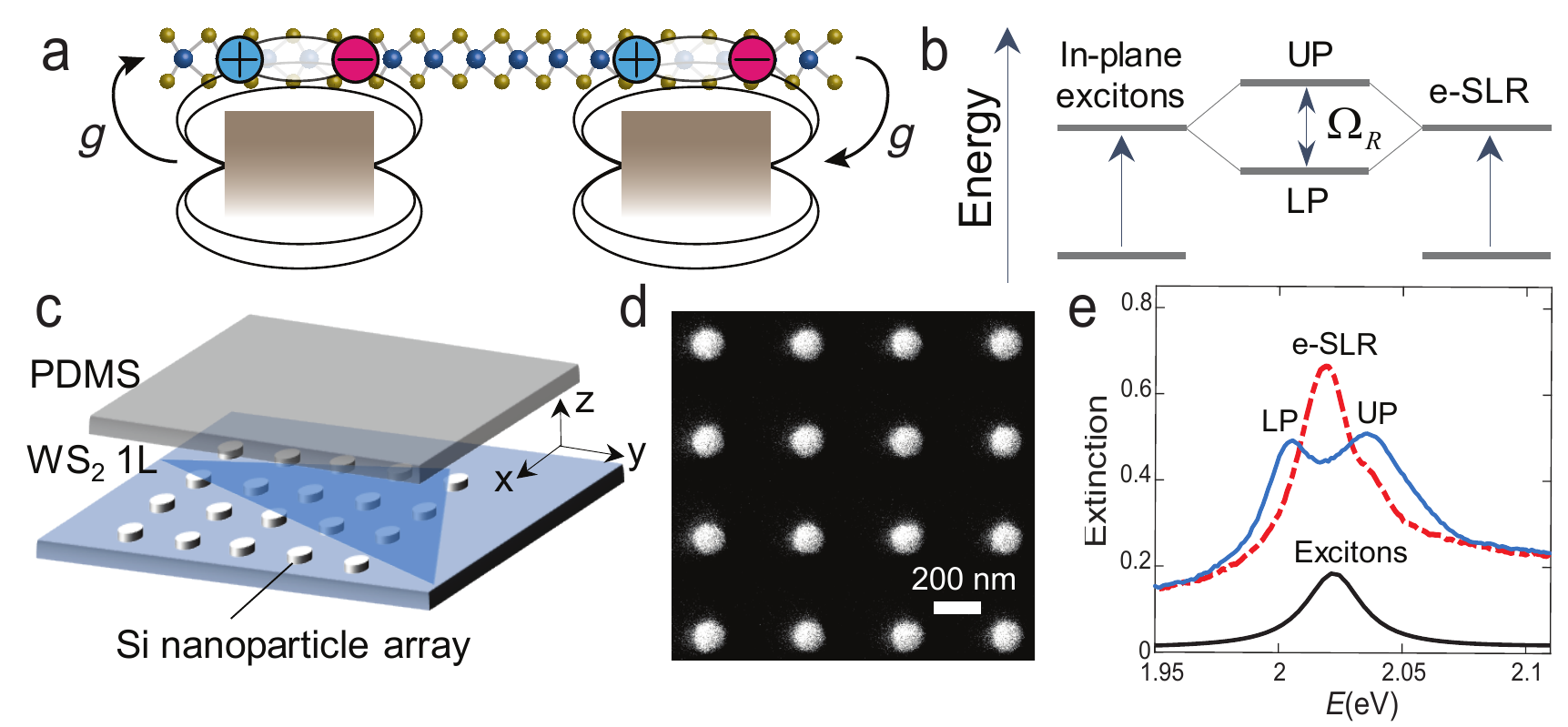}
\caption{Mechanism for in-plane coupling between dielectric nanoparticles and 2D semiconductors. (a) Illustration of a side view of two nanoparticles in an array with electromagnetic field coupling to in-plane excitons in a monolayer semiconductor on top. (b) Coherent energy exchange between electric dipole surface lattice resonances (one type of collective Mie-Resonances: e-SLR) in Si nanodisk arrays and bright A excitons of monolayer WS$_2$, leading to the formation of the upper (UP) and lower polariton (LP) states. The energy difference between the UP and LP states is the Rabi energy $\Omega_{R}$. (c) Schematic of the sample. A monolayer of WS$_2$  (WS$_2$~1L) is sandwiched between a superstrate of polydimethylsiloxane (PDMS) and the Si nanoparticle array on a fused quartz substrate. The WS$_2$~1L is in the $xy$ plane, while the out-of-plane direction corresponds to the $z$ axis. (d) Scanning electron microscope image (top view) of the Si nanoparticle array. (e) Extinction spectra of the bare e-SLR and excitons, and the coupled system. The strong coupling between e-SLR (red-dashed curve) and excitons (black-solid curve) leads to the splitting into two new peaks corresponding to UP and LP bands (blue solid curve).}
\label{fig:Figure01}
\end{figure}
\newpage                            

\begin{figure}
\includegraphics[width=0.8\textwidth]{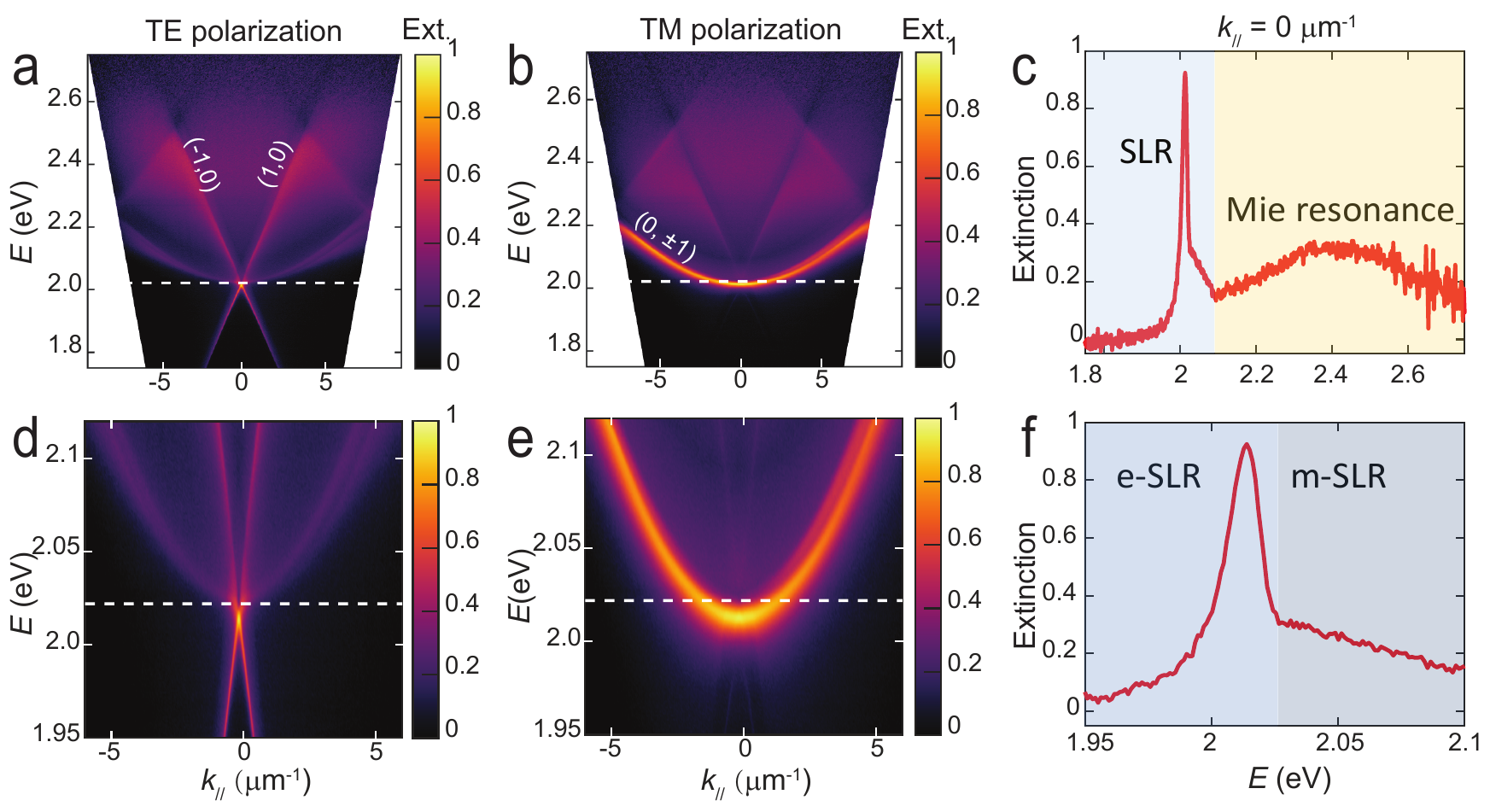}
\caption{Extinction dispersion of bare Si nanoparticle arrays. Panels (a)(d) and (b)(e) are recorded with TE and TM polarization, respectively. The dashed horizontal line represents the energy of bare A excitons of WS$_2$~1L. (c)(f) Extinction spectra for normal incidence, i.e., $k_\parallel=0$. The energy range in panels a-c allows to observe both the dispersive SLRs and the broader Mie-resonance. The extinction spectra of panels d-f are close-up views of the e-SLR and m-SLR. The degenerate mode at normal incidence splits into three different bands, including two linear dispersion bands associated with the (1, 0) and (-1, 0) diffraction orders and the parabolic band associated with the (0, $\pm 1$) order.}
\label{fig:Figure2}
\end{figure}
\newpage    

\begin{figure}
\includegraphics[width=0.8\textwidth]{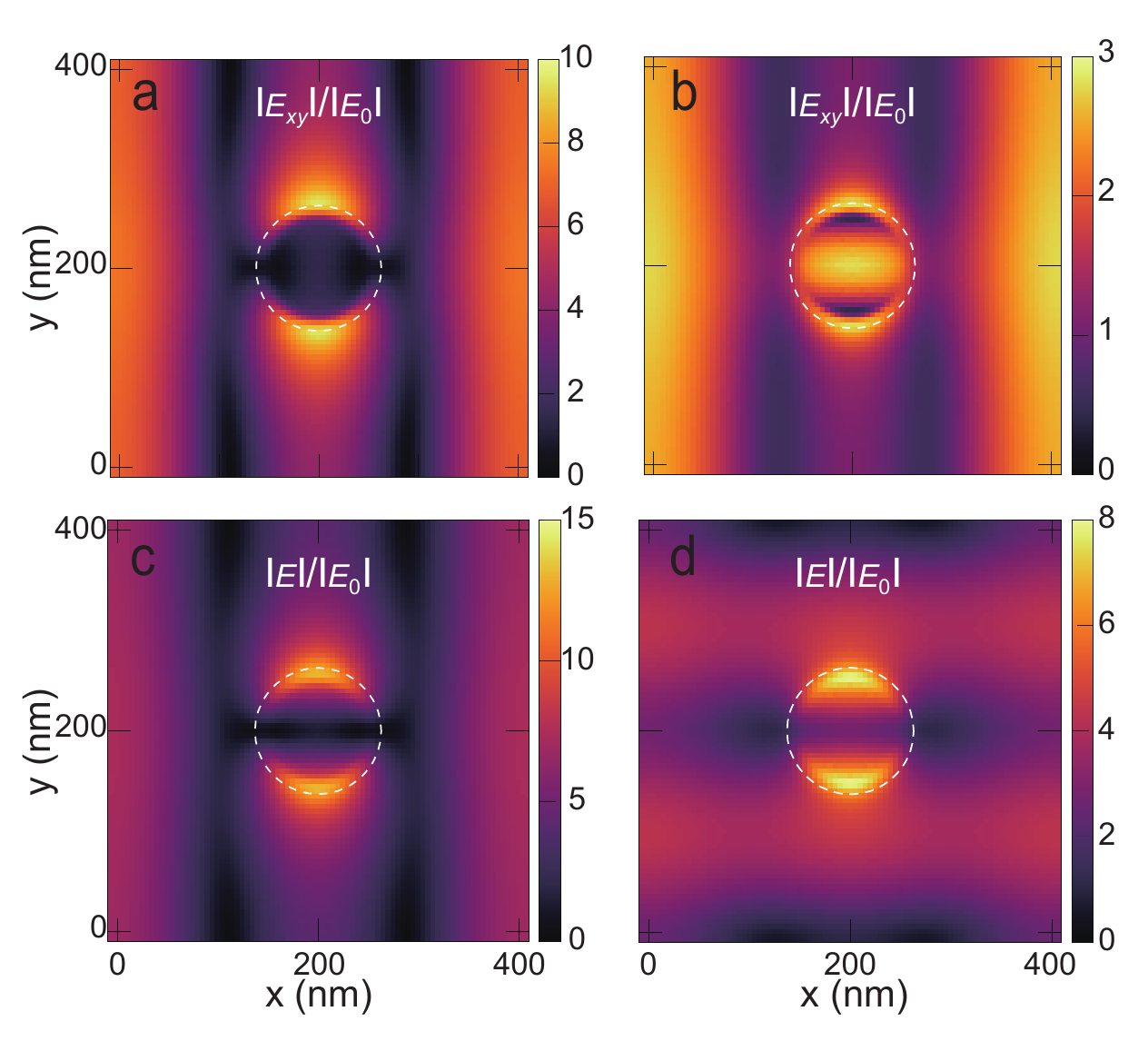}
\caption{
(a,b) Simulated in-plane electric field enhancement $\left | E{_{xy}} \right |/\left | E{_{0}} \right |$ and (c,d) total electric field enhancement $\left | E \right |/\left | E{_{0}} \right |$, in a unit cell of the Si particle array under normal incidence. The energy of the resonances in (a,c) corresponds to the e-SLR (2.006~eV) and (b,d) to the m-SLR (2.036~eV).}
\label{fig:Figure3}
\end{figure}
\newpage 

\begin{figure}
\includegraphics[width=0.8\textwidth]{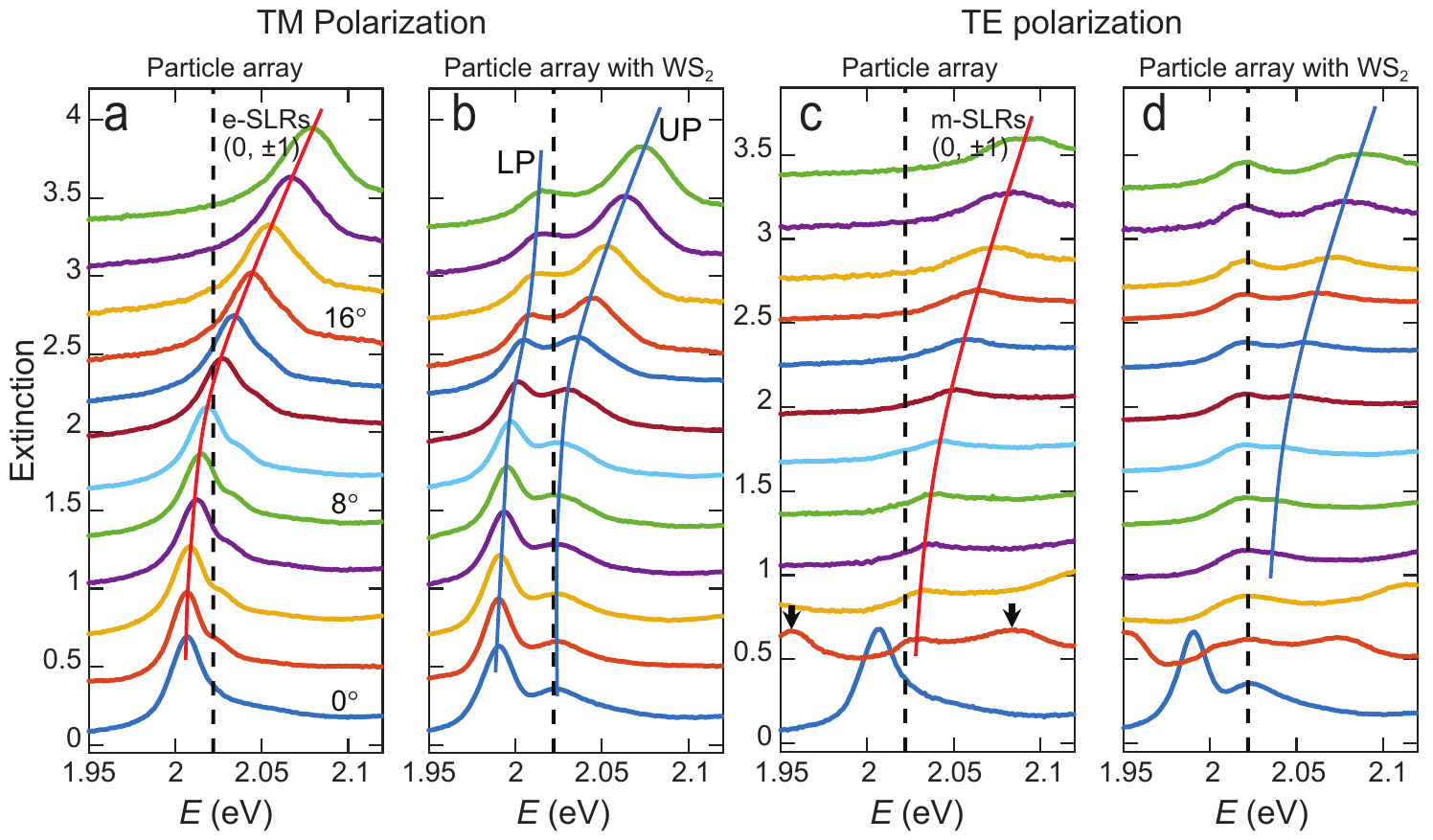}
\caption{Extinction spectra of the Si nanoparticle array without ((a) and (c)) and with ((b) and (d)) WS$_2$~1L on top as a function of the angle of incidence $\theta$ from $\theta=0^{\circ}$ to $22^{\circ}$. The optical micrograph (top view) of the measured sample is shown in Figure S1. Panels (a)(b) and (c)(d) are recorded with the TM and TE polarization, respectively. The extinction spectra in all the panels are offset for clarity. The black arrows in panel (c) indicate the peak energies of the ($\pm 1$, 0) e-SLRs. The red solid curves in (a) and (c) are guides to the eye illustrating the dispersion of the (0, $\pm 1$) e-SLRs and m-SLRs, respectively. The blue solid curves in panel (b) and (d) are guides to the eye illustrating coupled modes in the Si particle array and WS$_2$~1L system. The black vertical dashed lines represent the energy of A-excitons in WS$_2$~1L obtained from the extinction peak of the bare flake shown in Figure 1f.}     
\label{fig:Figure4}
\end{figure}
\newpage 

\begin{figure}
\includegraphics[width=0.8\textwidth]{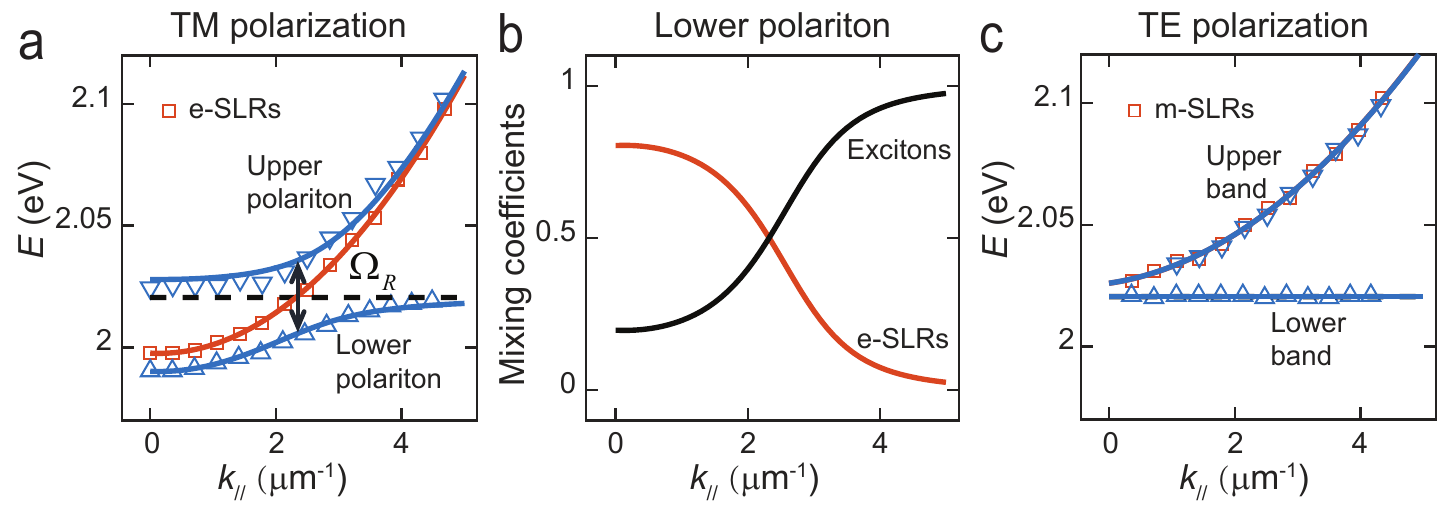}
\caption{The angular dispersion of the coupled system. (a) Upper polariton (blue downward triangles) and lower polariton (blue upward triangles)  bands for TM polarization extracted from the solid-blue guide to the eye in Figure 3b, are fitted by a model with two coupled harmonic oscillators. The fit indicates that the energies of uncoupled e-SLR (red squares) redshift by 10 meV compared to the bare (0, $\pm 1$) e-SLR in Figure 3a (the solid-red guide to the eye) and $\Omega_{R}$ = 32 meV. The dashed horizontal line denotes the energy of bare A excitons in WS$_2$~1L. (b) Weight fractions, i.e., mixing coefficients, of the lower polariton band as a function of $k_\parallel$ in terms of e-SLRs (red curve) or excitons (black curve). (c) Upper energy (blue downward triangles) and lower energy bands (blue upward triangles) for TE polarization extracted from the solid-blue guide to the eye and peaks close to the dashed-black line in Figure 3d, respectively. The coupled oscillator model fit (blue curves) indicates that the energies of uncoupled m-SLRs (red squares) only redshift 2 meV compared to the band of bare (0, $\pm 1$) m-SLRs in Figure 3c (the solid-red guide to the eye) and $\Omega_{R}$ = 0 meV.} 
\label{fig:Figure04}
\end{figure}
\newpage   

\begin{figure}
\includegraphics[width=0.8\textwidth]{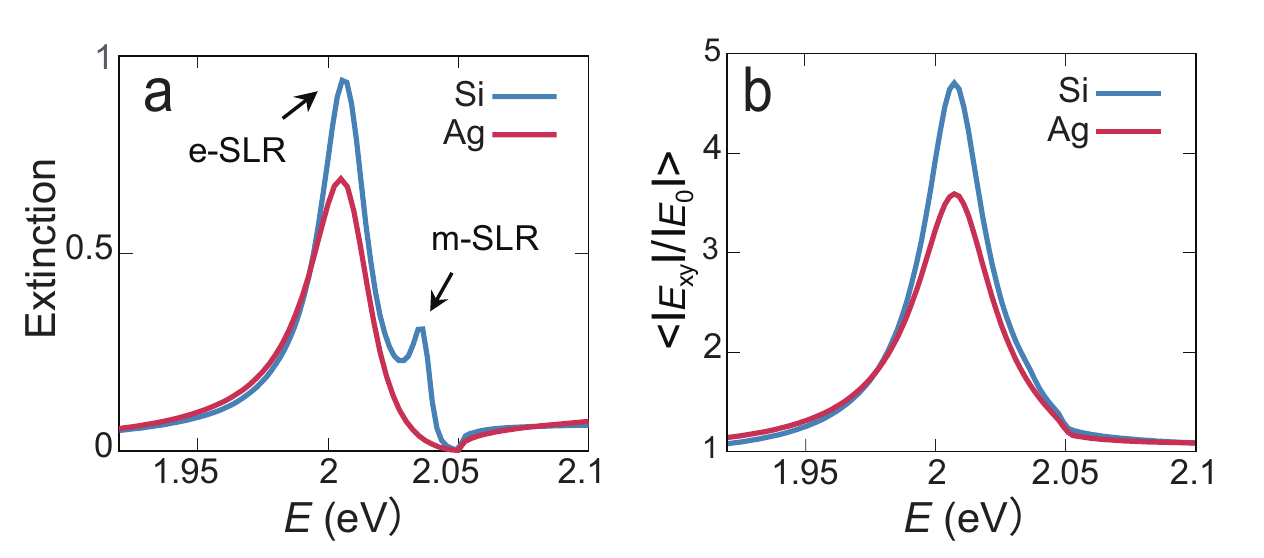}
\caption{Comparison of dielectric and metallic particle arrays for in-plane coupling. (a) The extinction spectra of Si particle array (blue curve) and Ag particle array (red curve) with the same square lattice and pitch of 420 nm. To tune the SLRs to the same resonance energy of the A exciton of WS$_2$~1L, the diameters of individual Ag/Si nanodisks are 74 nm and 126 nm, respectively, and the heights are 40 nm and 90 nm. The black arrows denote the peaks of e-SLR and m-SLR supported in Si particle array. (b) Root-mean-square in-plane field enhancement factor ($\left | E_{xy} \right |/\left | E{_{0}} \right |$) in a unit cell of the Si particle array (blue curve) and Ag particle array (red curve), as a function of energy.The simulated field enhancement of Si and Ag are shown in Figure 3a and Figure S8a, respectively.} 
\label{fig:Figure06}
\end{figure}
\newpage   

\begin{figure}
\includegraphics[width=0.8\textwidth]{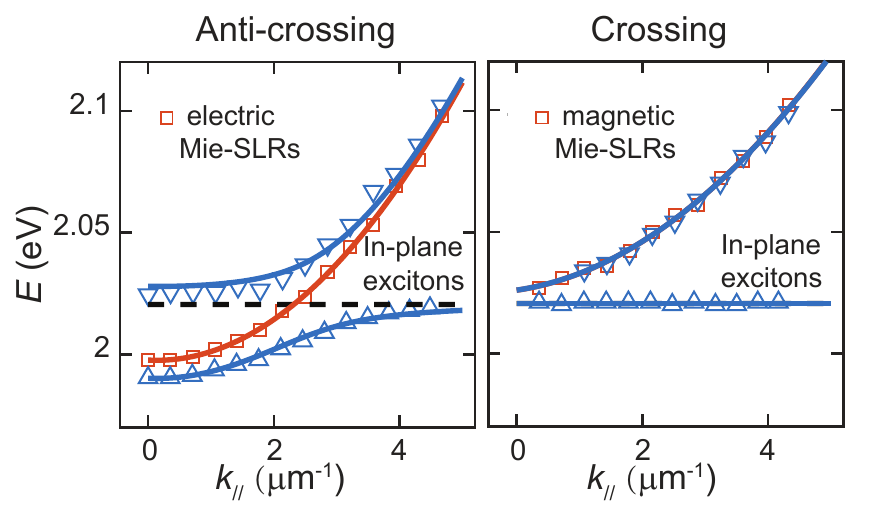}
\caption*{TOC Graphic}
\label{fig:TOC}
\end{figure}
\newpage       

\end{document}